\documentclass[journal]{IEEEtran}
\usepackage{graphicx}
\usepackage{amsmath}
\usepackage{textcomp}
\def\BibTeX{{\rm B\kern-.05em{\sc i\kern-.025em b}\kern-.08em
		T\kern-.1667em\lower.7ex\hbox{E}\kern-.125emX}}
\usepackage[normalem]{ulem}
\usepackage{amssymb,amsmath,bm}
\usepackage{subfigure}
\usepackage{amsfonts}
\usepackage{epsfig}
\usepackage{amssymb}
\usepackage{cite}
\usepackage[table,xcdraw]{xcolor}
\usepackage[utf8x]{inputenc}
\usepackage{color}
\usepackage{rotating}
\usepackage{tabularx}
\usepackage{array}
\usepackage{multirow}
\usepackage{blindtext}

\newtheorem{remark}{Remark}

\usepackage{moreverb}
\usepackage{epsfig}
\usepackage{adjustbox,lipsum}
\usepackage{epsfig}
\usepackage{anyfontsize}
\usepackage{epstopdf}
\usepackage[ruled,norelsize]{algorithm2e}
\makeatletter
\newcommand{\removelatexerror}{\let\@latex@error\@gobble}
\makeatother
\SetAlCapHSkip{0pt}
\makeatletter
\newcommand{\thealgorithm}{\arabic\algocf@float}

\newcommand{\AlgoCaptionFormat}{}

\renewcommand{\algocf@makecaption@ruled}[2]{%
	\global\sbox\algocf@capbox{\hskip\AlCapHSkip%
		\setlength{\hsize}{\columnwidth}
		\addtolength{\hsize}{-2\AlCapHSkip}
		\vtop{\AlgoCaptionFormat\algocf@captiontext{#1}{#2}}}
}
\makeatother
\usepackage{blindtext}
\usepackage{fancyhdr}
\usepackage[subfigure]{tocloft}
\usepackage[font={small}]{caption}
\usepackage{tcolorbox}
\usepackage{xcolor}
\definecolor{azure(colorwheel)}{rgb}{0.0, 0.5, 1.0}
\usepackage[colorlinks=true,allcolors=azure(colorwheel)]{hyperref}%
\allowdisplaybreaks
\usepackage{hyperref}
\begin{document}

\title{AoI Minimization in Energy Harvesting and Spectrum Sharing Enabled 6G Networks}

\author{Amir Hossein Zarif,~\IEEEmembership{Student Member,~IEEE,}
		\thanks{A. H. Zarif, P. Azmi, and N. Mokari are with the Department of Electrical and Computer Engineering, Tarbiat Modares University, Tehran 14115111, Iran (e-mail: a.zarif@modares.ac.ir; pazmi@modares.ac.ir; nader.mokari@modares.ac.ir).}%
        Paeiz Azmi,~\IEEEmembership{Senior Member,~IEEE,}
        Nader Mokari,~\IEEEmembership{Senior Member,~IEEE,}
        Mohammad Reza Javan,~\IEEEmembership{Senior Member,~IEEE}
        \thanks{M. R. Javan is with the Department of Electrical and Robotic Engineering, Shahrood University of Technology, Shahrood 3619995161, Iran (e-mail: javan@shahroodut.ac.ir).}%
        and~Eduard Jorswieck,~\IEEEmembership{Fellow,~IEEE}
        \thanks{E. A. Jorswieck is with the Department of Information Theory and Communication Systems, TU Braunschweig, Germany (e-mail: e.jorswieck@tu-bs.de).}%
    }


\maketitle

	\begin{abstract}
		Spectrum sharing is a method to solve the problem of frequency spectrum deficiency. This paper studies a novel AI based spectrum sharing and energy harvesting system in which the freshness of information (AoI) is guaranteed. The system includes a primary user with access rights to the spectrum and a secondary user. The secondary user is an energy harvesting sensor that intends to use the primary user’s spectrum opportunistically. The problem is formulated as partially observable Markov decision processes (POMDPs) and solved using two methods: a deep Q-network (DQN) and dueling double deep Q-Network (D3QN) to achieve the optimal policy. The purpose is to choose the best action adaptively in every time slot based on its situation in both overlay and underlay modes to minimize the average AoI of the secondary user. Finally, simulation experiments are performed to evaluate the effectiveness of the proposed scheme compared to the overlay mode. According to the results, the average AoI in the proposed system is less than that of the existing models, including only overlay mode. The average user access improved from 30\% in the overlay mode to 45\% in the DQN and 48\% in the D3QN.
	\end{abstract}


	\section{INTRODUCTION}
		\IEEEPARstart{I}{n} recent years, due to a severe increase in use of frequency bands, especially in the new sixth generation (6G) networks, which is called internet of everything (IoE), the frequency spectrum's optimal use is needed. Cognitive radio networks (CRNs) are useful and practical tools for solving the problem of coexistence between wireless systems. In wireless communications, the CRN forms a design pattern for the network, which can effectively change its transmission mode by preventing interference on the primary user (PU) from secondary user (SU) and can improve the use of spectrum. In CRN, SU can detect unused bands with spectrum sensing, temporarily occupy them and send information \cite{Multiband_1,Cooperative_2}. There are two overlay and underlay modes in these networks, which in the first case, the SU is allowed to send information only in the absence of the PU. However, in the second case, the SU can start sending information with limited transmit power simultaneously and without interfering on the PU. On the other hand, artificial intelligence (AI) plays a significant role in 6G and can be widely used in many fields \cite{Spectrum_3}. Furthermore, with the increase of wireless nodes, information's timely arrival is critical in time-sensitive cases. For example, in vehicle-to-vehicle networking or natural disaster monitoring, it is essential to update the status of the process.
		
		The information age concept is introduced to measure the novelty of information. Therefore, maintaining the freshness of information needs considering a new network design and scheme specially for CRN. \cite{Minimizing_5}. Age of information (AoI) is the amount of time elapsed since the most recent received packet.
		
		In the early work on AoI, the queuing theory's theoretical perspective made it possible to analyze and characterize the age \cite{Minimizing_8,Age_9,Age_10}. The authors in \cite{Minimizing_8} consider a wireless broadcast network in which a base station sends time-sensitive information to multiple clients, and \cite{Age_9} considers AoI in general multihop networks. In \cite{Age_10}, research on the AoI of PU in cognitive radio based on queuing theory is explored. In \cite{Age_11,Scheduling_12}, the problem is formulated as a Markov decision process (MDP) to find dynamic transmission scheduling schemes in wireless broadcast networks.
		
		Recently, AoI is studied in energy harvesting systems \cite{Age_13,Scheduling_14}. Energy harvesting is a promising method that can provide sustainable and long-lasting power for communication networks. In an energy harvesting communication system, nodes can obtain energy from their surrounding environment. For example, environmental energy sources can be radio frequency (RF) waves and infrared radiation \cite{Energy_15}. Due to the randomness of energy harvesting, if the necessary energy for updating is not available, the update fails, and the information in these systems may be out of date. Therefore, the goal is optimal energy management to keep the information fresh. In \cite{Age_13}, the authors consider AoI for energy harvesting system with offline and online solutions. They show that the offline solution can minimize the average AoI, and the online solution can be close to the offline with balance updating. In \cite{Optimal_17}, the authors consider an online status update to minimize the long term average AoI under energy constraint. They use a best-effort uniform status update, energy-aware adaptive status update, and threshold structure for infinite, finite, and one unit battery size, respectively. The authors in \cite{Sending_18} investigate the tradeoff between the message rate and average AoI under zero-wait, energy timing adaptive, and threshold-based transmission policy. The real-time status update is considered in \cite{Age_26}, where the SU can relay the status packets from the PU to the destination in CRN. The problem is formulated as a constrained Markov decision process (CMDP) to minimize average AoI. \cite{Age_27} investigates AoI in a system consisting of an access point and a smart device where the access point sends information and energy to the smart device over block fading channels. The smart device receives data and stores the harvested energy until collecting enough energy. After that, the smart device sends one block of transmission to the destination. \cite{Average_28} considers a sensor network where a sensor node harvests energy from a dedicated energy source to transmit status updates. The authors investigate the optimal value of the sensor battery to maximize the freshness of the information. The authors in \cite{Minimizing_29} assume a scheduling system consisting of physical sources and two sensors that observe status packets to minimize the AoI of the sources. However, they do not consider energy constraints for the system model. In \cite{Minimizing_30}, the authors consider a CRN consisting of an internet of things (IoT) device as a SU who aims to opportunistically access the PU spectrum. They formulate the problem as CMDP to investigate the minimization of long-term average AoI based on optimal policy under a collision constraint.
		
		In \cite{Optimal_20}, stochastic processes for primary spectrum availability are considered by modeling the problem as partially observable Markov decision processes (POMDP) subject to energy causality and collision constraints. The SU decides to remain idle or sense the spectrum to find a chance to occupy the spectrum and transmit. An optimal spectrum sensing policy is derived to maximize the expected total throughput subject to energy causality and collision constraints in \cite{Cognitive_21}. In \cite{Age_4}, an energy harvesting sensor as a SU sends status updates by opportunistically accessing the primary user’s spectrum in the overlay mode, i.e., when the spectrum remains idle. The problem is modeled as POMDP and solved by dynamic programming for finite and infinite horizon.
		
		All the mentioned cases consider modeling only the overlay mode. However, in cognitive radio, the environment is unknown, and we need to solve our problem in an online manner with energy management. Furthermore, AoI can be reduced by using both overlay and underlay modes, where the SU can access the spectrum with the full or limited power.
		
		In this paper, an energy harvesting and spectrum sharing system is modeled in both overlay and underlay modes by POMDP and solved by deep Q-network (DQN) and dueling double deep Q-Network(D3QN). The goal is to find an optimal action in every time slot that minimizes the AoI of the SU. The significant contributions and novelties of this paper can be summarized as follows:
		\begin{itemize}
			\item A spectrum sharing model is designed between the primary and secondary users in both overlay and underlay modes, in which the SU has an energy harvesting sensor. The SU can send data with full power in the overlay mode, or with the limited power in the underlay mode.
			\item The proposed system model is formulated as a partially observable Markov decision process (POMDP) to minimize AoI under casualty constraint.
			\item We solve the proposed POMDP problem with deep Q-network and dueling double deep Q-Network and compare their results to achieve better performance and stability. We need online, adaptive, and intelligent methods, especially in CRNs and 6G since the environment is unknown.
			\item Finally, the results show that the SU's access to the spectrum increases, and as a result, the AoI is reduced. In other words, the AoI in the proposed model is less than the baseline model, which only includes the overlay mode.
		\end{itemize}
		The remainder of the paper is organized as follows. We introduce the considered energy harvesting and spectrum sharing system in the overlay and underlay modes in Section II. In Section III, we formulate POMDP and specify its parameters. In Section IV, we propose the DQN based algorithm with experience reply and target network to solve this problem. In Section V, we solve our system model by D3QN to achieve better performance and quick convergence. In Section VI, the proposed power allocation and rate of the SU are expressed based on its situation relative to the PU. Section VII illustrates numerical results and Section VIII concludes the paper.

	\section{SYSTEM MODEL}
		We consider one PU, one energy harvesting SU, and a central entity for controlling power of SU in a CRN as shown in Fig. 1. The SU collects data and energy from its environment by a wireless sensor to provide real-time status updates, sends this data to its destination, and when its status updates are successful, the SU receives 1-bit feedback signal. We consider a time-slotted system with $t = 0,1,…,T-1$. The system model consisting of the model for PU and SU, is formulated as a Markov chain and POMDP, respectively. In Table I, the symbols and their meanings are listed for ease of reference.

		\begin{table}[!t]
			\renewcommand{\arraystretch}{1.3}
			\caption{Summary of Symbols}
			\label{Symbols}
			\centering
			\begin{tabular}{|c|c|}
				\hline
				\bfseries \textbf{Symbols} & \bfseries \textbf{Operational Meaning}\\
				\hline
				$A$ & Active state of PU\\
				\hline
				$I$ & Inactive state of PU\\
				\hline
				$N_{0}$ & Noise floor\\
				\hline
				$N_{\text{th}}$ & Acceptable threshold in underlay mode\\
				\hline
				$P_{r}$ & Revived power\\
				\hline
				$P_{t}$ & Transmit power\\
				\hline
				$o_{t}$ & State of PU\\
				\hline
				$\hat o_{t}$ & Observation of state of PU\\
				\hline
				$x_{t}$ & Action space of SU\\
				\hline
				$Z_{t}$ & Sensing action of SU\\
				\hline
				$L_{t}$ & Location of SU\\
				\hline
				$W_{t}$ & Wait period for receive power from center\\
				\hline
				$U_{t}$ & Status update of SU\\
				\hline
				$P_{f}$ & Probability of false alarm\\
				\hline
				$P_{d}$ & Probability of detection\\
				\hline
				$h_{t}$ & Channel gain\\
				\hline
				$e_{t}$ & Energy harvested\\
				\hline
				${\upsilon _1}$ & Average Poisson distribution in energy harvested\\
				\hline
				${\upsilon _2}$ & Average Normal distribution in energy harvested\\
				\hline
				$B_{\text{max}}$ & Battery capacity\\
				\hline
				$\alpha$ & Energy consumption for sensing\\
				\hline
				$\varphi ({h_t})$ & Energy consumption for updating\\
				\hline
				$\delta$ & Energy consumption for sending location\\
				\hline
				$b_{t}$ & Battery state\\
				\hline
				$a_{t}$ & AoI state\\
				\hline
				$s_{t}$ & State space\\
				\hline
				$P(.|.)$ & Probability of transition\\
				\hline
				$r_{t}$ & Immediate reward\\
				\hline
				$\pi$ & Policy\\
				\hline
				$V(s)$ & State-value function\\
				\hline
				$Q(s)$ & Action-value function\\
				\hline
				$\theta$ & Weight parameter set of the Q-network\\
				\hline
				${\theta '}$ & Weight parameter set updated in every N steps\\
				\hline
				$\gamma$ & discount factor\\
				\hline
				$\varepsilon$ & Greedy policy\\
				\hline
				$L(\theta)$ & Loss function\\
				\hline
				$\beta$ & Learning rate\\
				\hline
				$D$ & Distance between PU and SU\\
				\hline
				$\omega$ & Path loss exponent\\
				\hline
				$\lambda$ & Wavelength\\
				\hline
				$c$ & Speed of light\\
				\hline
				$f$ & Frequency\\
				\hline
			\end{tabular}
		\end{table}

	\subsection{PU Model}
		The PU has the right to access the spectrum entirely and arbitrarily, so in any time slot, it can be active (A), i.e., transmitting its information, or inactive (I) which forms a Markov chain with the state of $o_t\in \{A, I\}$. The two-state (active or inactive) Markov chain model is commonly used for modeling PU activity \cite{Primary_23}. Markov chain for the PU has the transition probabilities $P_{\text{II}}$, $P_{\text{AI}}$, $P_{\text{AA}}$ and $P_{\text{IA}}$, which are defined as staying in inactive state, transitioning from A to I, staying in active state, and transitioning from I to A respectively.
		
		\begin{figure}[!t]
			\centering
			\includegraphics[width=2.5in]{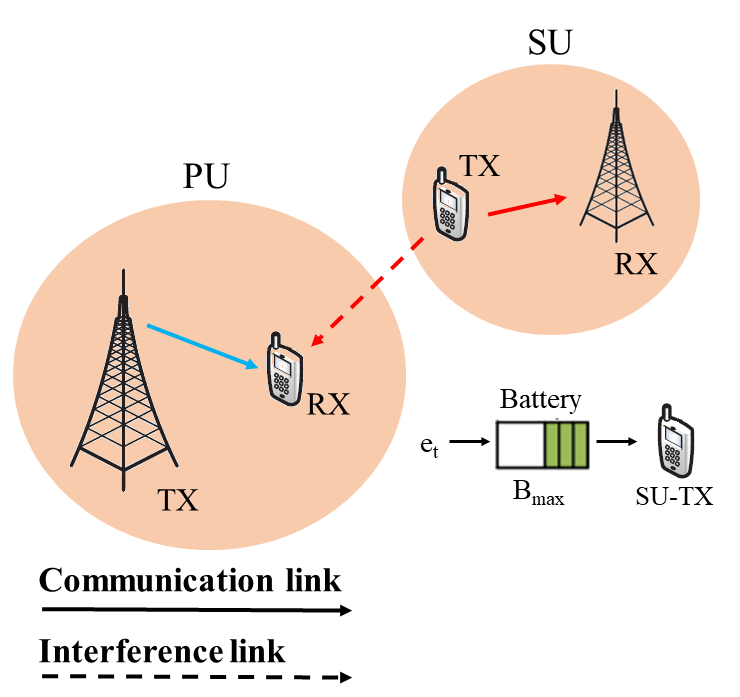}
			\caption{The system model. The SU can sense and access
				opportunistically the PU’s spectrum.}
			\label{System Model}
		\end{figure}

	\subsection{SU Model}
		At the beginning of each time slot, SU decides to sense the channel or not, i.e., $Z_t$. If the sensing result shows that the PU is inactive, i.e., the received power in the SU is less than the specified threshold, the SU may decide whether to send the status $U_t$ or not in the overlay mode. On the other hand, if the result shows the PU is active but the received power in the SU is less than the threshold $N_{\text{th}}$, the SU has to send its location $L_t$ to the central entity which it has no interference and waits to receive the appropriate power from it $W_t$, and as a result, the SU can send its status in the underlay mode. Furthermore, suppose the PU is active and the received power in the SU is more than the threshold. In that case, the SU has to be silent because the SU is within the PU coverage, and according to the spectrum sharing rules, the PU must be protected from any interference. We assume all steps occur in one-time slot because the SU has an energy harvesting wireless sensor and its status and location are not big data.
		Let $x_t$=$(Z_t,L_t,W_t,U_t) \in X$ denote the decision tuple, i.e., actions the SU can choose them in every slot $t$, where
		\begin{displaymath}
			Z_t\in\{0~(\text{idle}),1~(\text{sense})\},
		\end{displaymath}  
		\begin{displaymath}
			L_t\in\{0~(\text{not~send~location}),1~(\text{send~location})\},
		\end{displaymath} 
		\begin{displaymath}
			W_t\in\{0~(\text{not~wait}),1~(\text{wait})\},
		\end{displaymath} 
		\begin{displaymath}
			U_t\in\{0~(\text{not~update}),1~(\text{update})\}.
		\end{displaymath}                       
		Therefore, the acceptable actions for each time slot that the SU can choose are $X$=$\{(0,0,0,0),(1,0,0,0),(1,0,0,1),(1,1,1,1),(1,1,1,0)\}$. Action $(0,0,0,0)$ is equal to no sense where the SU decides not to do anything because it does not have enough energy. Action $(1,0,0,0)$ is the silent mode where the PU is active, and the SU cannot send data or the SU intends to send the location in the underlay mode but does not have enough energy. Action $(1,0,0,1)$ means overlay mode and action $(1,1,1,1)$ is the underlay mode. Also, action $(1,1,1,0)$ means the underlay mode but the power received from the central entity is equal to zero, which means the SU is wrong in sensing or the SU does not have enough energy to send the information.
		
		The decisions are made adaptively based on SU’s states and its knowledge of the primary spectrum availability as introduced below:
	\subsubsection{Observation Model}
		The SU opportunistically accesses the spectrum by sensing the environment. The observation can be obtained based on SU's action and state, that is, its experience. Hence, an observation of the state of PU is denoted by $\hat o_t$=$\{I,A~and~Pr<N_{\text{th}},A~and~Pr>N_{\text{th}}\}$. We consider imperfect sensing where false alarm, i.e., detecting PU is active when it is not, and miss detection, i.e., detecting PU is inactive when it is transmitting, events can occur. Let $P_f$ denote the probability of false alarm, and $P_d$ denote the probability of detection defined by
		\begin{equation}
			{P_f} = \text{Pr} ({\hat o_{t}} = A|{o_t} = I)= \text{Pr} ({P_r} > {N_0}|{o_t} = I),
		\end{equation}
		\begin{equation}
			{P_d} = \text{Pr} ({\hat o_{t}} = A|{o_t} = A)= \text{Pr} ({P_r} > {N_0}|{o_t} = A),
		\end{equation}
		where $N_0$ is the average noise power.
	\subsubsection{Channel Model}
		The SU transmits data over block fading channel with channel gain $h_t$ for slot $t$ where $\left| {{h_t}} \right|$ is random variable follows a Rayleigh distribution, and it is independently and identically distributed (i.i.d) over slots.
	
	\subsubsection{Energy Harvesting Model}
		The SU can harvest energy from the environment and store it in its battery. The energy harvested at slot $t$ is $e_t$ and can be a discrete random variable with the Poisson distribution with mean ${\upsilon _1}$ or continuous random variable with positive Normal distribution with mean ${\upsilon _2}$ and variance ${\sigma _2^2}$ when ${\upsilon _2}>>{\sigma _2}$ and they are i.i.d across slots as
		\begin{equation}
				\text{Discrete}:~{\rm{Pr}}({e_t} = k) = \exp \{  - {\upsilon _1}\} \frac{{{{({\upsilon _1})}^k}}}{{k!}},
		\end{equation}
		\begin{equation}
				\text{Continuous}:~\scalebox{0.89} {$f({e_t}) = \left\{ \begin{array}{l}
					\frac{1}{{{\sigma _2}\sqrt {2\pi } }}\exp \{  - \frac{1}{2}{(\frac{{{e_t} - {\upsilon _2}}}{{{\sigma _2}}})^2}\} ~~~~~~ {\upsilon _2} >> {\sigma _2}\;\\
					0 \hspace{4.25cm} \text{otherwise}.
				\end{array} \right.$}
		\end{equation}

		In other words, in model simulation and analysis, we will use both distributions for the harvested energy and compare the results. The harvested energy is used to sense the primary spectrum, send the location to the central, and update the status over the wireless channel. Let $\alpha$ be the energy consumption of sensing. The transmission energy depends on the channel inversion, so we denote the energy cost for an update in slot $t$ by $\varphi ({h_t})$ where the function  is non-increasing, and the updating cost is $\varphi ({h_t}) = \frac{1}{{{h_t}}}$ \cite{Joint_24}. We assume that energy consumption for sending location to the center is $\delta$. Because the packet size of the sending location is very small compared to the original data, we assume that the amount of energy consumed is constant \cite{Joint_24}. The battery state in each time slot is $b_t = \{0,..., B_{\text{max}}\}$, battery capacity is $B_{\text{max}}$ and $b_0$ correspond to the initial energy stored in the battery where
		\begin{equation}
			b_{t + 1} = \min \{ {b_t} + {e_t} - \alpha {Z_t} - \varphi ({h_t}){U_t} - \delta {L_t},{B_{\max }}\}.
		\end{equation}
		One of the primary considerations in energy harvesting communication systems is that the SU cannot use the harvested energy immediately at the same time. Hence, the energy consumption causality constraint is given by
		\begin{equation}
			\alpha {Z_t} + \varphi ({h_t}){U_t} + \delta {L_t} \le {b_t}.
		\end{equation}
		If the SU cannot send its status, it can sense to update its observation about the environment. Note that the harvested energy is stored in the battery and then used at the next slot.
	\subsubsection{AoI Model}
		AoI measures the amount of time elapsed since the most recent received packet. Let $a_t$ denote AoI of slot $t$. Whenever the SU decides to update its status, it generates and sends a data packet. The data packet is small enough to generate and send instantaneously when the action, i.e., spectrum sensing and sending location, is completed. Hence, it spends one slot to receive an update. If the status update is successful, then the AoI is equal to 1 otherwise it is increased by 1. The AoI state is $a_t = \{1,2,..., a_{\text{max}}\}$, where $a_{\text{max}}$ indicates that information at given destination is expired. The update decision is
		\begin{equation}
			{a_{t + 1}} = \left\{ \begin{array}{l}
			1,~~~~~~~~~\text{if}~{x_t} = (1,1,1,1)~\text{or}~(1,0,0,1),~\text{ACK}\\
			{a_t} + 1,~~~\text{otherwise}.
			\end{array} \right.
		\end{equation}
		Also, we assume the channel is error free so that the data can be received successfully. However, if the PU is active, $P_r>N_{\text{th}}$ and the SU state is $x_t = (1,0,0,1)$, i.e., miss detection, there is a collision that leads to an update failure.

	\subsubsection{Coexistence modes of PU and SU}
		In this paper, the SU aims to access to primary spectrum in overlay and underlay modes opportunistically. If the SU decides to sense, there are several cases which are shown in Fig. 2. as follows:
		\begin{itemize}
			\item The PU is absent so that the SU can access the spectrum with full power, and it does not require to send location.
			\item The PU is present, but $N_0 < P_r < N_{\text{th}}$, where $N_0$ is the average noise power. The value of $N_{\text{th}}$ is chosen so that it is neither too close to $N_0$ nor too far away from it because if it is close to the noise power, the problem of non-sending occurs despite having the right position. If the distance is increased, it causes much interference for PU receiver. Therefore, its value is selected as an initial start for SU access to the spectrum. Hence, the SU can access the spectrum with limited power. In this case, the SU has to send its location to the central entity to receive appropriate power. Note that if the SU receives power equal to 0 from the central entity, it means that the distance between the PU receiver and SU transmitter is small, and if the SU starts to transmit, the interference created for the PU is greater than the allowable value. This event may be due to the effect of channel conditions, i.e., fading or hidden node problem, on the SU's received signal. We assume the central knows the location of the PU in each time slot. Since the SU has energy limitations and the spectrum sensing is not done ideally with complete accuracy, sending the location to the central entity causes less energy consumption and more PU protection.
			\item The PU is present and $P_r > N_{\text{th}}$. Therefore, the SU cannot send anything and must be silent.
		\end{itemize}

		\begin{figure}[!t]
			\centering
			\includegraphics[width=3.2in]{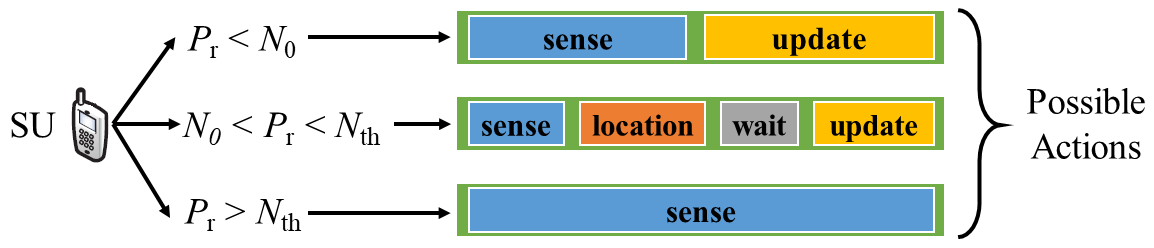}
			\caption{Coexistence modes of PU and SU with possible actions for the SU when having the enough energy.}
			\label{Scenarios}
		\end{figure}
	
		\begin{remark}
			Note that because in addition to protecting the PU from interference, the privacy of the PU must also be protected in spectrum sharing, the SU cannot have PU's location. On the other hand, in underlay mode, transmission with limited power must be done to maintain interference. Therefore, in this case, the SU needs to get help from the central entity to follow spectrum sharing principles. In this case, the errors due to spectrum sensing are significantly reduced. In other words, the SU strategy is semi-decentralized.
		\end{remark}

	\section{POMDP FORMULATION}
		\subsection{Observation state space}
			In this section, we formulate the POMDP to determine an optimal action. The main goal of this work is to minimize AoI of the SU in the spectrum sharing system for both overlay and underlay modes. When the SU decides an action in every time slot, one of the following observations may occur from its environment.
		\subsubsection{Case 1}
			If the SU stays idle without sensing due to insufficient energy based on its decision, it does not have any new observation.
		\subsubsection{Case 2}
			If the SU decides to sense the channel and finds the PU is inactive, the SU transmits in the overlay mode and successfully updates its status by receiving a feedback signal.
		\subsubsection{Case 3}
			If the SU senses the channel is empty, and the SU fails to update its status, a collision has occurred.
		\subsubsection{Case 4}
			The PU is sensed to be inactive and the SU decides not to update because the energy required to send is more than the stored energy.
		\subsubsection{Case 5}
			If the SU senses the PU is active and $P_r > N_{\text{th}}$, the SU does not update to prevent interference for the PU.
		\subsubsection{Case 6}
			If the SU sensed the PU is active but $P_r < N_{\text{th}}$, the SU transmits in the underlay mode and a receives feedback signal. Hence, the SU successfully updates its status.
		\subsubsection{Case 7}
			If the SU sensed the PU is active and $P_r < N_{\text{th}}$, but the SU does not have required energy, the SU decides not to update.
		\subsubsection{Case 8}
			The PU is sensed to be active and $P_r < N_{\text{th}}$, but the SU received power equal to zero from the central entity for updating the status. Hence the SU decides not to update because it makes a mistake in spectrum sensing, i.e., due to the effect of channel conditions like fading.
			
		 The agent can observe only partial information of the environment states. Therefore, we determine the state based on current observations and information, i.e., $s_t = (a_t, b_t, e_t, {[{P_r}]_t})$. If the SU does not perform the spectrum sensing, then ${P_r} = 0$.
			
		\subsection{Transition probabilities}
			The transition probability of SU to next state $s_{t+1} = (a_{t+1}, b_{t+1}, e_{t+1}, {[{P_r}]_{t+1}})$ given current state $s_t = (a_t, b_t, e_t, {[{P_r}]_t})$ and action $x_t = (Z_t, L_t, W_t, U_t)$ is denoted by ${P}(s_{t+1} | s_t,{x_t})$. Since the energy harvesting and the channel fading are i.i.d, the transition probabilities is equal to
			\begin{equation}
				\begin{split}{P}({s_{t + 1}}|{s_t},{x_t}) = P({a_{t + 1}}|{a_t},{x_t})P({b_{t + 1}}|{b_t},{e_t},{x_t}) \\
				\times P({e_{t + 1}})P({h_{t + 1}}), \end{split}
			\end{equation}
			where
			\begin{equation}
				{P({a_{t + 1}}|{a_t},{x_t})} = \left\{ \begin{array}{l}
				1, ~~  \text{if}~{a_{t + 1}} = (1 - {U_t}){a_t} + 1\\
				0, ~~  \text{otherwise},
				\end{array} \right.
			\end{equation}

			\begin{equation}
				\scalebox{0.9} {$P({b_{t + 1}}|{b_t},{e_t},{x_t}) = \left\{ \begin{array}{l}
				1,~~\text{if}~{b_{t + 1}} = \min \{ {B_{\max }},\\
				~~~~~~~~{b_t} + {e_t} - \alpha {Z_t} - \varphi ({h_t}){U_t} - \delta {L_t}\} \\
				0,~~\text{otherwise}.
				\end{array} \right.$}
			\end{equation}
		\subsection{Reward Function}
			Let $r_t$ be the immediate reward under the influence of current state $s_t$ and action $x_t$. Therefore, we have
			\begin{equation}
				\scalebox{0.9} {${r_t} = \xi ({R_{\text{SU}}}) - {a_t} = \left\{ \begin{array}{l}
					- {a_t}, \hspace{2.28cm} \text{no~sensing}\\
					- {a_t}, \hspace{2.3cm} \text{silent}\\
					\xi {R_{\text{full}}} - {a_t}, \hspace{1.35cm} \text{overlay}\\
					\xi {R_{\text{limited}}} - {a_t}, \hspace{1cm} \text{underlay}
				\end{array} \right.$}
			\end{equation}
			where $\xi$ is intended to balance the reward and ${R_{\text{SU}}}$ is rate of SU which is equal to $R_{\text{full}}$ and $R_{\text{limited}}$ in overlay and underlay modes respectively which is described in Section VI. In other words, the reward consists of some state parameters to improve learning and convergence of the network.
			\begin{remark}
				Note that by maximizing the reward $r_t$, the SU increases its rate while trying to reduce the AoI. Since the best mode of access for the SU is the overlay mode, where the SU can use the spectrum freely and with maximum efficiency due to the lack of the PU.
			\end{remark}
		\subsection{Policy}
			Consider a sequence of functions that will be referred to as the policy $\pi=\{\mu_0,..., \mu_{T-1}\}$ to find a decision rule. $\mu_t$ is a mapping from a state of the system ${s_t}$ into an action ${x_t}$. The reward is generally a random variable, therefore formulated as expected reward. The goal is to find the optimal sensing and update policy that maximize reward function or equivalently minimize AoI. In other words, the policy must be changed so that it can achieve optimal action in every slot to minimize AoI.

	\section{DEEP Q NETWORK}
		When the state space and action space is small, the Q-learning algorithm can effectively obtain the optimal policy. However, with an uncertain spectrum environment and no prior knowledge, i.e., the state transition probability is unknown, especially in CRN, the belief space is large, and the Q-learning algorithm may not find the optimal policy. Thus, to solve POMDP, we use deep Q-Network (DQN) algorithm  to approximate the action-value function \cite{Deep_25}, i.e.,
		$Q({s_t},{x_t};\theta ) \approx {Q^*}({s_t},{x_t})$ where $\theta$ is the weight parameter set of the Q-network.
		
		\begin{figure}[!t]
			\label{Algorithm1}
			\removelatexerror
			\begin{algorithm}[H]
				\SetAlgoLined
				\caption{Deep Q-Network with experience reply and target network}
				\textbf{Input}: memory size $D$, mini-batch size $B$, discount rate $\gamma$, learning rate $\beta$, $\varepsilon$ in $\varepsilon$-greedy policy and the number of iterations $I_{\text{max}} > 0$.\\
				Initialize the Q-network $Q({s_t},{x_t};\theta )$ and its target network $Q({s_t},{x_t};\hat \theta )$ with random weights.\\
				Initialize the starting action $x_0$ and execute it to get the initial state $s_0$.\\
				Initialize Train=\textbf{True}\\
				\For{$t=0,1,...$}{
					\eIf{Train}{
							Choose $x_t$ by $\varepsilon$-greedy policy.
						}{
							${x_{t + 1}} = \arg \max  Q({s_t},{x_t};\theta )$
						} 
					Apply action $x_t$ and collect $r_t$ and $s_t$.\\
					\If{Train}{
						Store $(s_t, x_{t+1}, r_{t+1}, s_{t+1})$ in memory unit.\\
						\If{$t \ge D$}{
							Remove the oldest experience tuple in memory unit.
					    }
						Sample random mini-batch of experience tuples from memory unit.\\
						Compute gradient descent on the loss function ${L_t}(\theta )$ and update the weights $\theta$.\\
						Every $N$ steps copy the weights $\theta  \to \theta '$.\\
						\If{$t-D > I_{\text{max}}$}{ 
							Train=\textbf{False}
					    }
					}
				}
			\end{algorithm}
		\end{figure}
		
		DQN is a reinforcement learning method that combines Q-learning with convolutional neural network (CNN). Let ${V_\pi }(s)$ denote state-value function under policy $\pi$
		\begin{equation}
			{V_\pi }(s) = \mathop {\max }\limits_{x \in X} \;\; \mathbb{E} \left\{ \sum\limits_{t = 0}^{T - 1} {{\gamma ^t}r_t|{s_t},{x_t}} \right\},
		\end{equation}
		which represents the maximum expected accumulated reward starting given $({s_t},{x_t})$ and depends on the policy. In other words, the state-value function indicates being valuable of that state during previous experiences. Let ${Q_\pi }({s_t},{x_t};\theta )$ denote the action-value function or Q-function, which is the maximum expected accumulated reward starting from $t$ for taking an action ${x_t \in X}$ under policy ${\pi}$ and consists of the immediate reward obtained from the current state and the expected sum of value functions for the next slot. Therefore, the POMDP based on Bellman equation can be solved as follows:
		\begin{equation}
			{V_\pi }(s) = \mathop {\max }\limits_{x \in X} \;\;{Q_\pi }({s_t},{x_t};\theta ),
		\end{equation}
		and the goal is to find the optimal action selection policy as
		\begin{equation}
			{\pi ^*} = \mathop {\arg \max }\limits_{x \in X}  {Q_\pi }({s_t},{x_t};\theta ).
		\end{equation}

		The main process of DQN to find minimum AoI is detailed in Algorithm 1. Each component of DQN is specified as follows:
		\subsubsection{Input Layer}
			The inputs of DQN are the state $s_t$ including AoI, amount of energy harvested, battery state at time slot $t$ and the corresponding observation, i.e., ${s_t} = ({a_t},{b_t},{e_t},{[{P_r}]_{t}})$.
		\subsubsection{Output Layer}
			The output of DQN is estimated Q-value, i.e, $Q({s_t},{x_t};\theta )$.
		\subsubsection{Reward Definition}
			The reward $r_t$ after taking action $x_t$ is defined in (11) and the cumulative rewards of each time slot is formulated as follows:
			\begin{equation}
				R' = \sum\limits_{t = 0}^\infty  {{\gamma ^t}{r_t}},
			\end{equation}
			where $\gamma$ is the discount factor and $\gamma \in (0,1)$.
		\subsubsection{Q-Network}
			The Q-network maps the current state to a series of action values, i.e., $Q({s_t},{x_t};\theta )$. Once $\theta$ is learned, the Q-values can be determined, and the action with the largest Q-value will be taken in each time slot.
		\subsubsection{Target Network}
			Let us define the target network as $Q({s_t},{x_t};\theta ')$. The target network is similar to Q-network structure, but the main network is updated in every time slot, while the target network is updated every N steps. In other words, the target network gets weights of Q-network every N steps to ensures that the Q-network is received stable values in the learning process.
		\subsubsection{Action Selection}
			At first, by setting the initial stage, each state-action pair's Q-value is not correct because the network has not converged. If the action with the largest Q-value is selected, most actions will not be executed, and the corresponding Q-values will not be updated effectively. Therefore, to make a trade-off between exploitation and exploration, $\varepsilon$-greedy policy is adopted for action selection, where the possibility of action ${x_{t + 1}} = \arg \max  Q({s_t},{x_t};\theta )$ is $1-\varepsilon$, and the possibility of random action is $\varepsilon$. $\varepsilon  \in (0,1)$ is a small positive value.
		\subsubsection{Experience Reply}
			An experience tuple $(s_t, x_{t+1}, r_{t+1}, s_{t+1})$ is stored in a memory, and a mini-batch of experience tuples will be sampled in each iteration for training the network.
		\subsubsection{Loss Function}
			The loss function to update the weight parameter set $\theta$, called temporal difference (TD) error, is defined as follows:
			\begin{equation}
				\begin{aligned}
				{L_t}(\theta ) = {\mathbb{E}_{{s_t},{x_t},{s_{t + 1}},r}}{\{ [r_t + \gamma \mathop {\max }\limits_{x \in X} \;\;Q({s_{t + 1}},{x_{t + 1}};\theta ')} \\ {- Q({s_t},{x_t};\theta )] ^2\} },
				\end{aligned}	
			\end{equation}
			where the symbol ${\theta '}$ is only updated with $\theta$ in every N steps from the same Q-network. Differentiating the loss function w.r.t the weights, we arrive at the following gradient:
			\begin{equation}
				\begin{aligned}
				{\nabla _\theta }{L_t}(\theta ) = {\mathbb{E}_{{s_t},{x_t},{s_{t + 1}},r}}\{ [r_t + \gamma \mathop {\max }\limits_{x \in X} \;\;Q({s_{t + 1}},{x_{t + 1}};\theta ') \\
			 	- Q({s_t},{x_t};\theta )]{\nabla _\theta }Q({s_t},{x_t};\theta )\}.
				\end{aligned}
			\end{equation}
			Based on each component of DQN, the Q-Network is updated as follows:
			\begin{equation}
				\begin{array}{l}
					Q({s_t},{x_t};\theta ) \leftarrow Q({s_t},{x_t};\theta ) + \beta [{r_t}\\
					+ \gamma \mathop {\max }\limits_{x \in X} Q({s_{t + 1}},{x_{t + 1}};\theta ') - Q({s_t},{x_t};\theta )],
				\end{array}
			\end{equation}
			where $\beta$ is the learning rate or step size, and it has a small value.

	\section{Dueling Double Deep Q Network}
		In this section, we implement the proposed model with Dueling Double Deep Q Network(D3QN). The structure of D3QN is generally similar to DQN, but some changes have been made to enhance the performance and prevent overestimation of action values. The reason for overestimation is the positive bias in updating the maximum function in Q-learning. Therefore, the concept of D3QN based on double DQN and dueling DQN is introduced \cite{D3QN_32}. The double DQN can effectively mitigate the overestimation and improves the performance of learning. It is expressed in the following equation:
		\begin{equation}
			\begin{array}{l}
				Q({s_t},{x_t};\theta ) \leftarrow Q({s_t},{x_t};\theta ) + \beta [{r_t}\\
				+ \gamma Q({s_{t + 1}},\mathop {\arg \max }\limits_x (Q({s_{t + 1}},{x_{t + 1}};\theta ));\theta ') - Q({s_t},{x_t};\theta )].
			\end{array}
		\end{equation}
	
		The dueling DQN consists of two streams: state value, $V$, and advantage function, $A$, which is combined in the end, producing a single Q-function. The output of the state value is a scalar, but the advantage function is an advantage vector with dimension of the number of actions. The advantage function, $A$, indicates how an action is necessary to the benefit function of all actions. Therefore, the Q value is the sum of the value $V$ and the advantage function $A$, which is expressed as
		\begin{equation}
			Q({s_t},{x_t};\theta ) = V({s_t};\theta ) + A({s_t},{x_t};\theta ) - \frac{1}{{|A|}}\sum\limits_{{x_{t + 1}}} {A({s_t},{x_{t + 1}};\theta )}.
		\end{equation}
		The main process of D3QN is like DQN, but the loss function for updating the weight parameter is defined as follows:
		\begin{equation}
			\scalebox{0.89}
			{$\begin{array}{l}
				{L_t}(\theta ) = {\mathbb{E}_{{s_t},{x_t},{s_{t + 1}},r}}\{ [{r_t} + \gamma Q({s_{t + 1}},\mathop {\arg \max }\limits_x (Q({s_{t + 1}},{x_{t + 1}};\theta ));\theta ')\\
				- Q({s_t},{x_t};\theta )]^2{\} }.
			\end{array}$}
		\end{equation}

	\section{POWER ALLOCATION AND THROUGHPUT}
		According to each of the following cases, the amount of power that the SU needs to send its status is specified after receiving the location of users.
		\begin{itemize}
			\item If the PU is not present in the channel, the SU starts sending update at full power, i.e., ${P_{\text{SU}}} = {P_{\text{full}}}$ with rate
				\begin{equation}
						{R_1} = \text{lo}{\text{g}_2}\left(1 + \frac{{{P_{\text{full}}} \times {{\left| {{h_t}} \right|}^2}}}{{\sigma _n^2}}\right).
				\end{equation}
			\item If the PU is present in the channel and $P_r>N_{\text{th}}$, the SU does not meet the permissible conditions. Hence, its power and rate are zero, $P_{\text{SU}}=R_2=0$.
			\item If the PU is present in the channel and $P_r<N_{\text{th}}$, the SU can send an update with the limited power and rate
				\begin{equation}
						{R_3} = \text{lo}{\text{g}_2}\left(1 + \frac{{{P_{\text{limited}}} \times {{\left| {{h_t}} \right|}^2}}}{{\sigma _n^2}}\right),
				\end{equation}
			where ${\sigma _n^2}$ is the average of noise.
		\end{itemize}
	
		\begin{table}[!t]
			\renewcommand{\arraystretch}{1.3}
			\caption{Hyper-parameters of DQN and D3QN}
			\label{Hyper-parameters}
			\centering
			\begin{tabular}{|c|c|}
				\hline
				\bfseries \textbf{Hyper-parameters} & \bfseries \textbf{Value}\\
				\hline
				Memory size $D$ & 2000\\
				\hline
				Mini-batch size $B$ & 32\\
				\hline
				Target network update $N$ & 35\\
				\hline
				Discount rate $\gamma$ & 0.95\\
				\hline
				Learning rate $\beta$ & 0.01\\
				\hline
				$\varepsilon$-greedy policy & $1 \to 0.001$\\
				\hline
				Activation function & ReLU\\
				\hline
				Optimizer & Adam\\
				\hline
			\end{tabular}
		\end{table}
	
		In the last case, the SU must send its location to the central entity and wait to receive an appropriate transmit power. Suppose the spatial position of PU is equal to $(x_1, y_1)$ and the spatial position of SU is equal to $(x_2, y_2)$. In this case, the distance between users is calculated as follows:
		\begin{equation}
			D = \sqrt {{{({x_1} - {x_2})}^2} + {{({y_1} - {y_2})}^2}}.
		\end{equation}
		Since PU always starts sending with the full power, using the Path Loss model to the PU center, we calculate the coverage radius as follows:
		\begin{equation}
			\frac{{{P_r}}}{{{P_t}}}~\text{dB} = 10{\log _{10}}k - 10\omega {\log _{10}}\frac{d}{{{d_0}}} - {\psi _{\text{dB}}},
		\end{equation}
		\begin{equation}
			k = 20{\log _{10}}\frac{\lambda }{{4\pi {d_0}}}~\text{[dB]},~~~\lambda  = \frac{c}{f},
		\end{equation}
		where $k$ is a constant factor, $\omega$ is the path loss exponent, $d_0$ is the reference distance, $\lambda$ is the wavelength, $f$ is frequency, and $c$ is speed of light. In addition, ${\psi _{\text{dB}}}$ is a Gauss-distributed random variable whose mean is zero and variance is $\sigma _{{\psi _{\text{dB}}}}^2$.
		The value of $P_t$ is equal to the transmitted power of PU, and the value of $P_r$ is equal to $N_{0}$. Therefore, the approximate distance obtained is equal to $d_{\text{th}}$ based on the considered path loss and average shadowing. That is, the SU must start sending information with the limited power outside it to prevent interference for PU. Because the sensing operation may not occur ideally without a false alarm, if the SU is placed in this area after sending its location and checking the distance, i.e., $d' = D - {d_{\text{th}}}<0$, the power assigned to it will be zero. To calculate the power assigned to SU, similar to the PU, the value of $P_r$ and the distance $D$ are known, and therefore the value of $P_t$ is obtained.

		\begin{figure}[!t]
			\centering
			\includegraphics[width=2.9in]{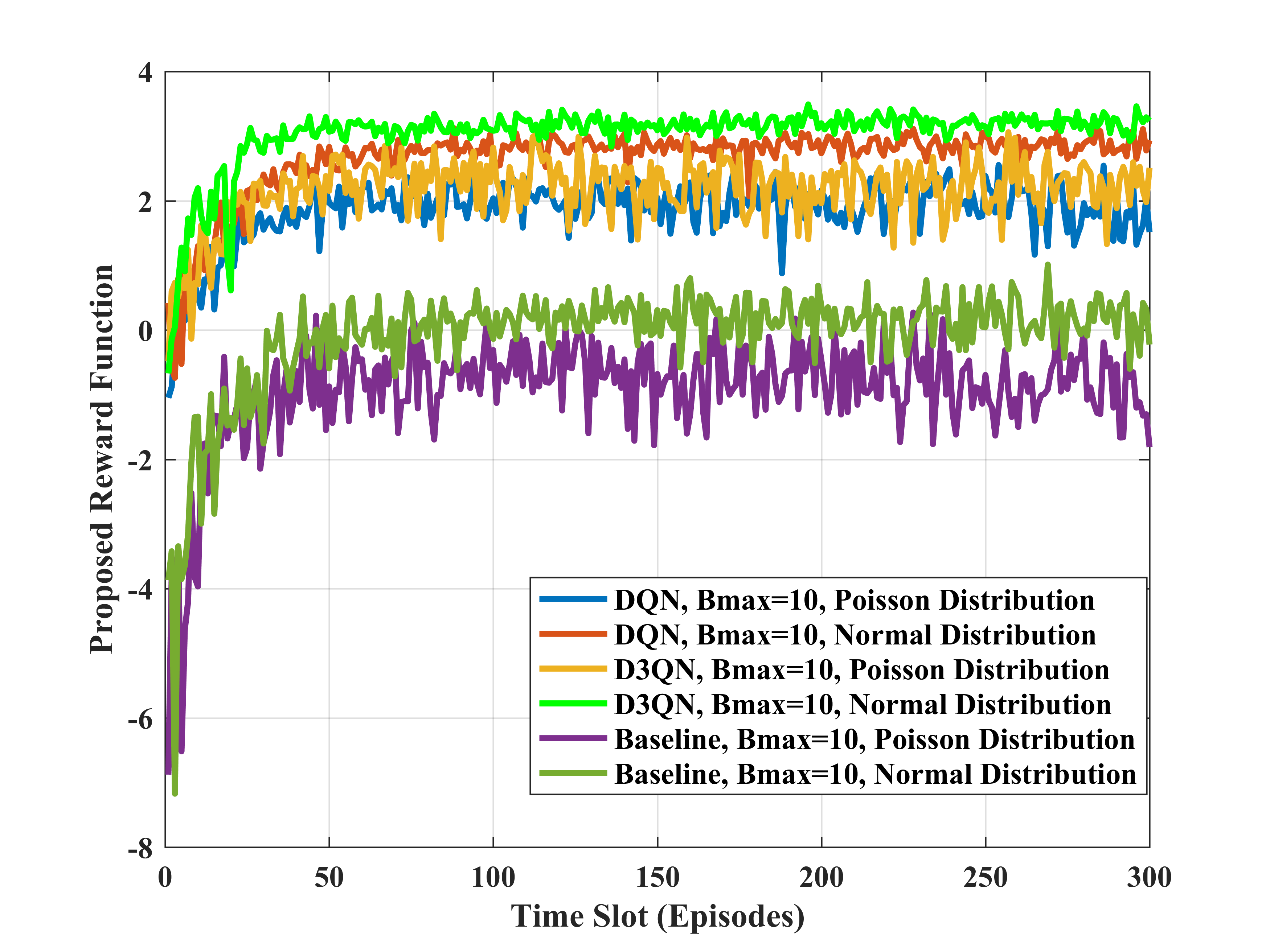}
			\caption{Reward function in the proposed and baseline schemes.}
			\label{Reward2}
		\end{figure}
		\begin{figure}[!t]
			\centering
			\includegraphics[width=2.9in]{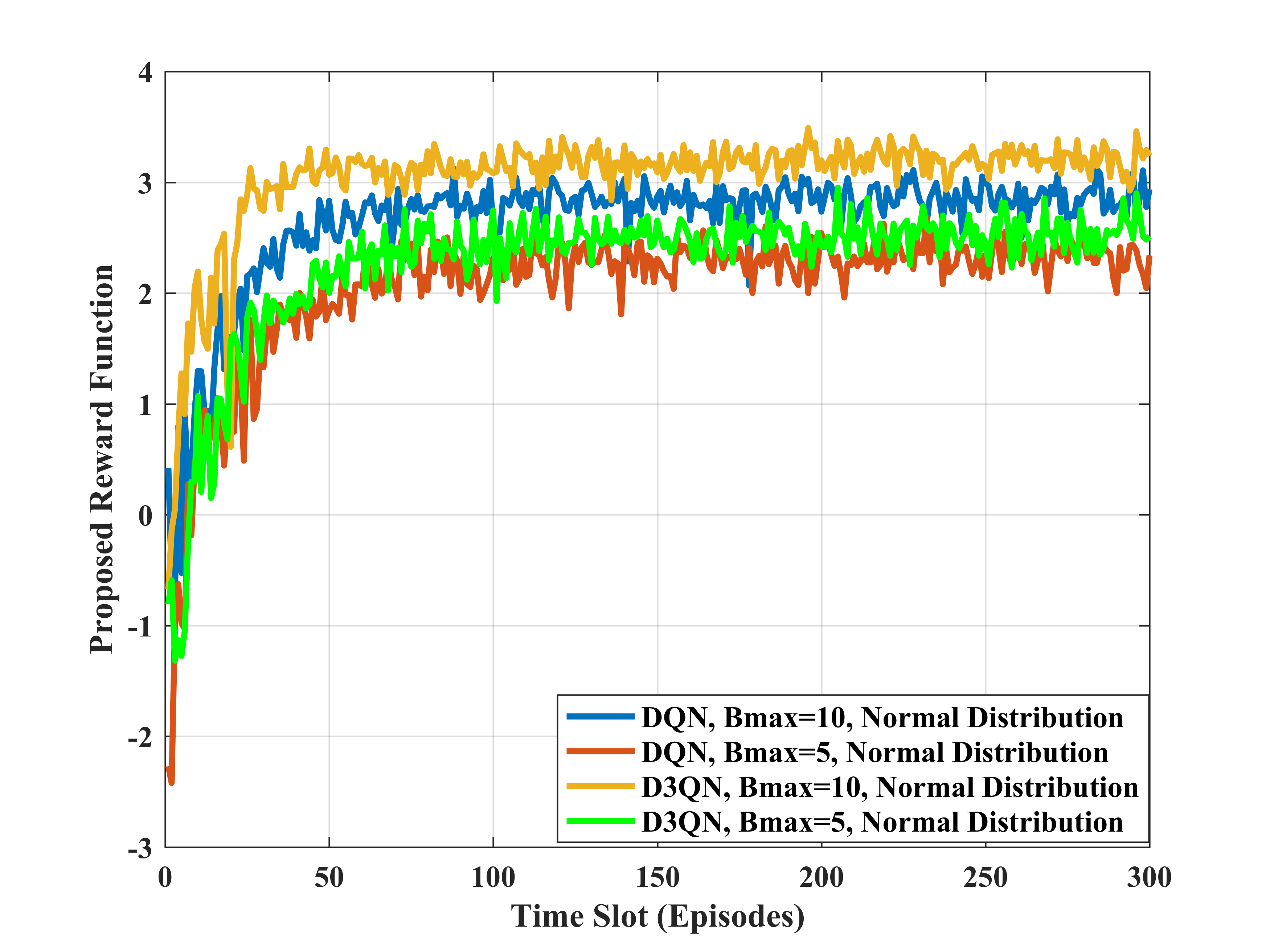}
			\caption{Reward function in the proposed scheme with $B_\text{max}=10,5$ mAH.}
			\label{Reward1}
		\end{figure}
	
	\section{SIMULATION RESULTS} 
		In this section, we present simulation results of our proposed partially observable Markov decision process with DQN and D3QN algorithms \footnote{The simulation code of this paper is available on: \url{https://dx.doi.org/10.21227/aynt-my59}}. Moreover, we compare the performance of our algorithms with a system that only includes the overlay mode. The neural network trained in DQN and D3QN have three fully connected layers consisting of two hidden layers containing 64 neurons and Rectified Linear Unit (ReLU) as activation function. The Adam algorithm \cite{adam_31} is used to perform Stochastic Gradient Descent to update the parameters of DQN and D3QN. Mini-batch size set 32, which is randomly selected from a memory size of 2000 prior experiences while updating the network's weights. $\varepsilon$ is initially set to 1 and is multiplied by 0.99986 every time step until reaching 0.001. The discount rate $\gamma$ is 0.95, and learning rate $\beta$ is 0.01. The number of time slots of the proposed schemes is 300, which iterate 200 times in every time slot and is implemented in Pytorch. Finally, the target network updates every 35 time slots. Hyper-parameters of the DQN and D3QN are summarized in Table II.

		In the simulations, the considered frequency is \hspace{2cm} $f=2.4~$GHz, maximum battery size of SU is $B_{\text{max}}=10~\text{or}~5$ mAH, energy consumption for spectrum sensing is $\alpha=3$, energy consumption for sending location to the center in the underlay mode is $\delta=1$, and the energy consumption for updating status is equal to inverse of channel gain. Furthermore, the noise power is $\sigma _n^2= -80$~dB, which is equal to $N_0$, the threshold between the underlay and silent mode is $N_\text{th}=-60$~dB, and the path loss exponent is $\omega=3$. Average Poisson and Normal distributions in the energy harvested system are ${\upsilon _1} = {\upsilon _2}=3$, the variance of Normal distribution is ${\sigma ^2}=0.25$, variance of shadow fading is $\sigma _{{\psi _{\text{dB}}}}^2 = 6~$dB and speed of light is $C=3 \times {10^8}~$m/s.

		\begin{figure}[!t]
			\centering
			\includegraphics[width=3in]{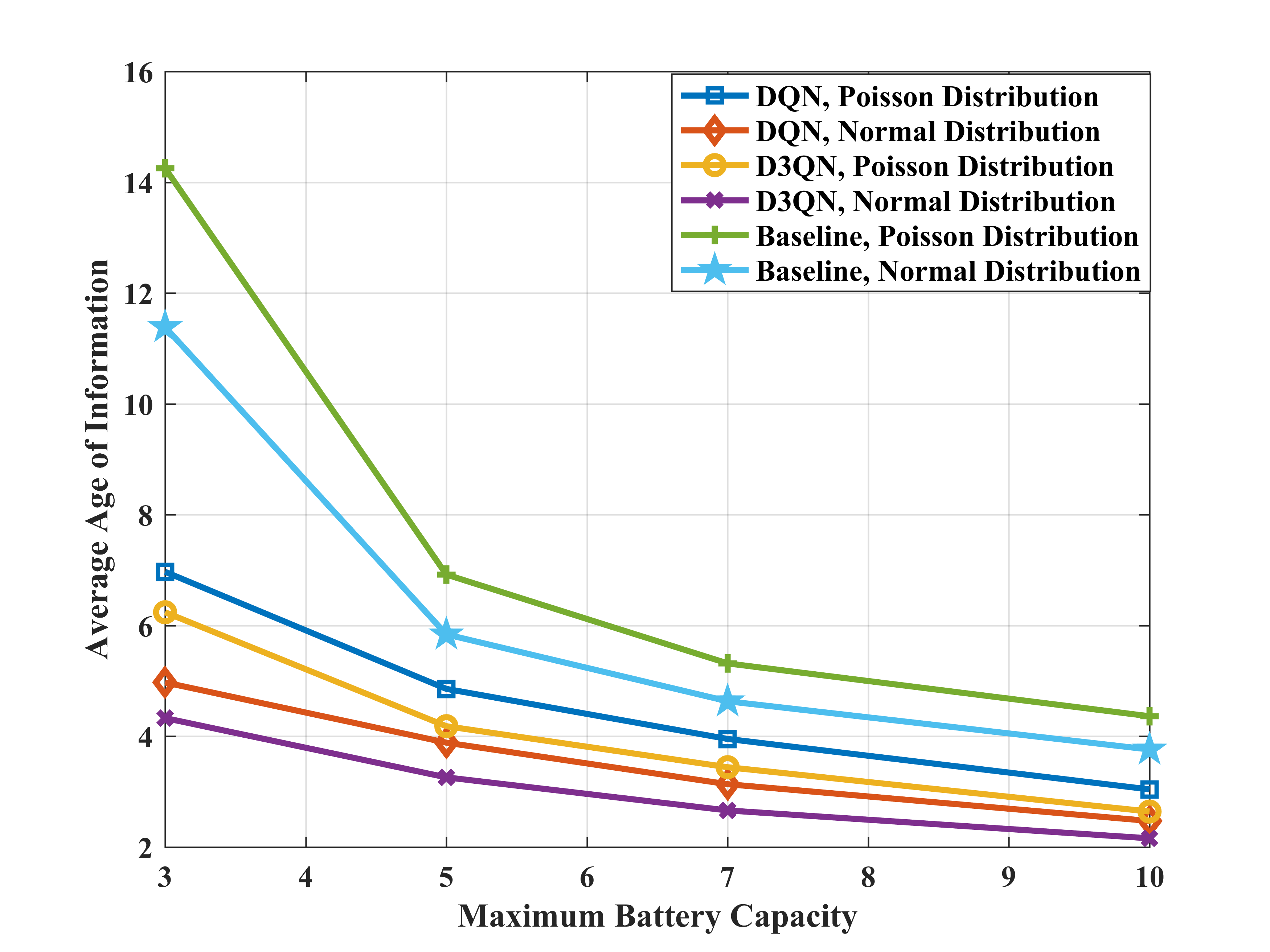}
			\caption{Average AoI vs maximum battery capacity in the proposed and baseline schemes.}
			\label{AoI1}
		\end{figure}

		In the results, the proposed model refers to the introduced system consisting of the overlay and underlay modes and the baseline model refers to the system consisting of only overlay mode. Figs. 3 and 4 show the evaluation and convergence of artificial intelligence used for different methods. In Fig. 3, the proposed reward function for different energy distributions, i.e., Poisson and Normal distributions, is plotted at the same battery capacity. Due to changing the amount of AoI in each method, the proposed case's reward is more than the baseline case. Fig. 4 shows the reward function for different energy distributions with two different battery capacities. Accordingly, changes in battery capacity affect AoI. Besides, Normal distribution is a more appropriate and accurate model for energy distribution in reality due to its continuity compared to Poisson distribution. Also, the reward of D3QN is slightly better than DQN and it has converged more quickly.

		\begin{figure}[!t]
			\centering
			\includegraphics[width=3in]{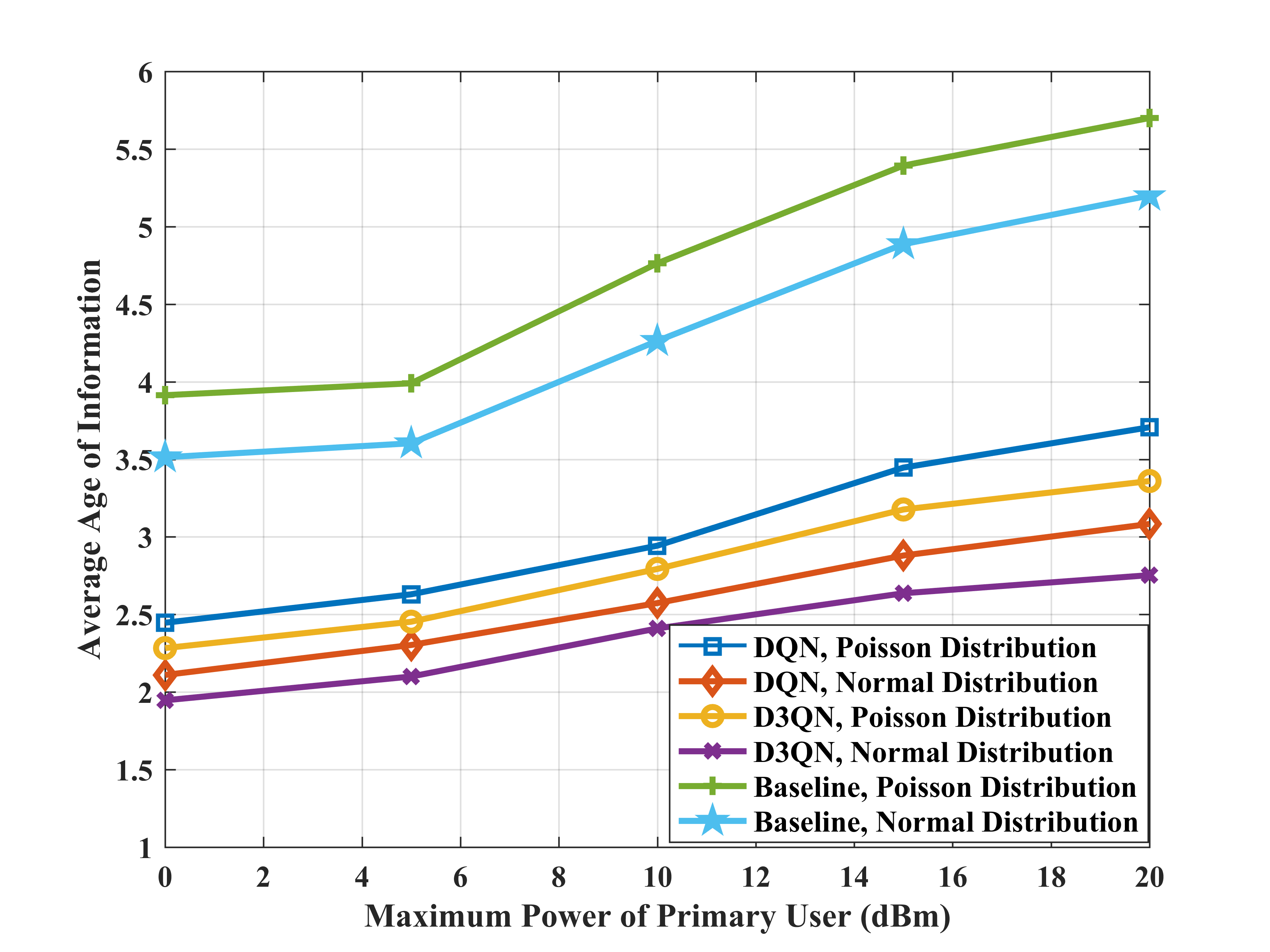}
			\caption{Average AoI vs maximum power of PU in the proposed and baseline schemes.}
			\label{AoI2}
		\end{figure}
		\begin{figure}[!t]
			\centering
			\includegraphics[width=3in]{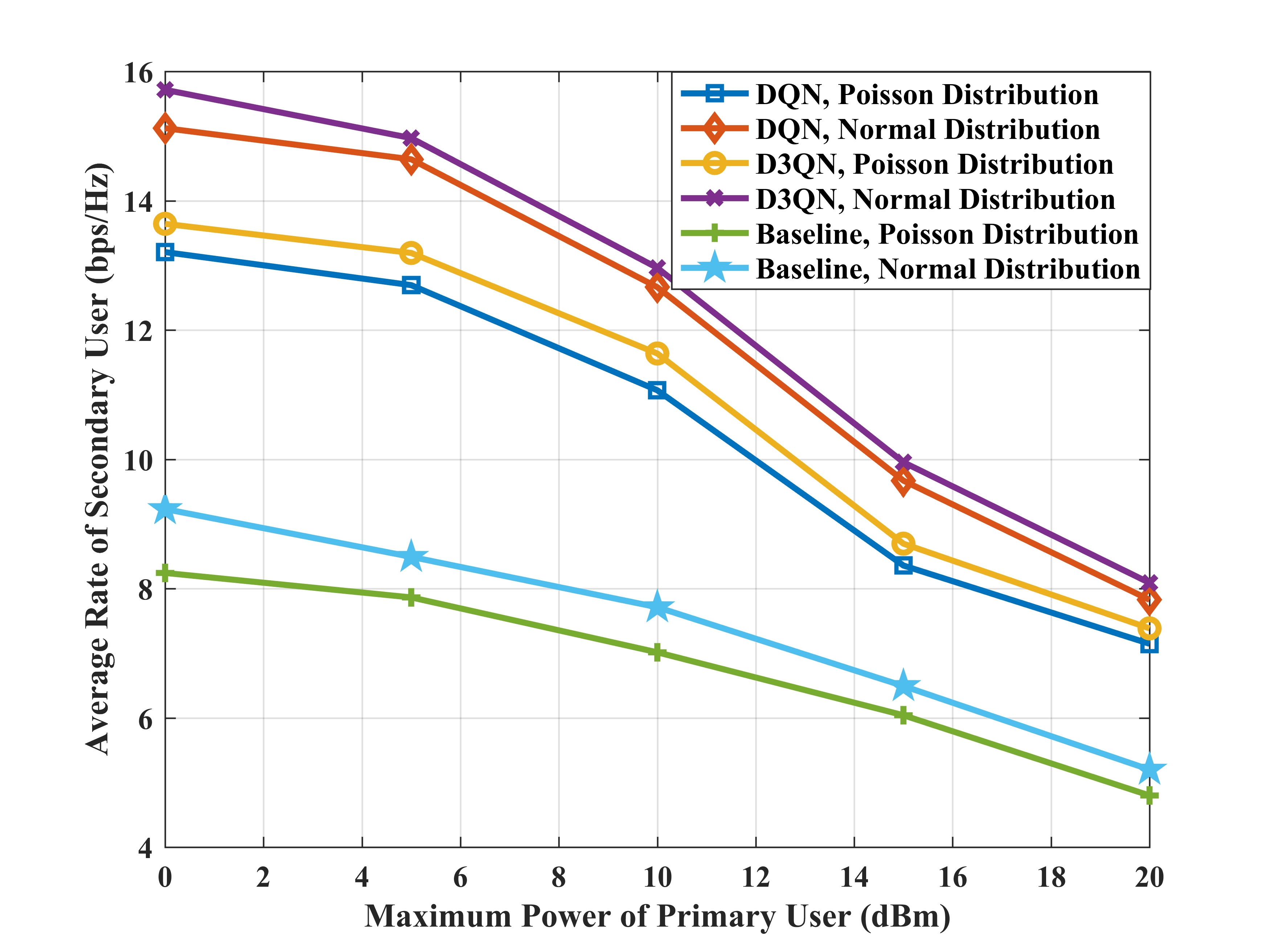}
			\caption{Average rate of SU vs maximum transmit power of PU in the proposed and baseline schemes.}
			\label{Rate}
		\end{figure}

		The average AoI based on different battery capacities is plotted in Fig. 5. As can be seen, increasing the battery capacity reduces the AoI. However, it is noteworthy that at lower capacities, the proposed method has much less AoI than the baseline method. The SU starts sending data with its maximum power in the baseline when the PU is absent based on the channel conditions, but in the proposed, the SU can select the underlay mode and send the data with the limited power if the received power is less than $N_\text{th}$. Also, the probability of the energy condition not being met is reduced by increasing the battery capacity. Fig. 6 shows the average AoI in terms of maximum power of PU. When the PU increases the amount of transmission power, the received power in the SU increases. In other words, the probability that the received power is higher than the defined threshold increases. Accordingly, in the baseline case, when power of the PU increases, the probability of the silent mode is higher than the overlay mode. However, in the proposed case, when the power increases, the SU has a chance to send data in the underlay mode with the limited power. Hence, the SU has more access to the channel, which reduces the AoI. Therefore, the average AoI in the proposed case is less than the baseline. Besides, the average AoI in the D3QN is less than the DQN due to better performance.
	
		\begin{table}[!t]
			\centering
			\caption{Comparison of DQN and D3QN with the baseline}
			\label{Table3}
			\begin{tabular}{|c|c|c|c|c|}
				\hline
				\multirow{2}{*}{Power of PU(dBm)} & \multicolumn{2}{c|}{Rate(bps/Hz)} & \multicolumn{2}{c|}{AoI} \\ \cline{2-5} 
				& DQN             & D3QN            & DQN         & D3QN       \\ \hline
				0                                 & +54.6\%         & +58.5\%         & -34.2\%     & -38.5\%    \\ \hline
				5                                 & +55.5\%         & +61.6\%         & -32.7\%     & -36.4\%    \\ \hline
				10                                & +51.4\%         & +59.2\%         & -35.5\%     & -38.8\%    \\ \hline
				15                                & +29.6\%         & +32.3\%         & -33.7\%     & -38.9\%    \\ \hline
				20                                & +31.9\%         & +35.7\%         & -32.7\%     & -37.8\%    \\ \hline
			\end{tabular}
		\end{table}
		\begin{figure}[!t]
			\centering
			\includegraphics[width=2.9in]{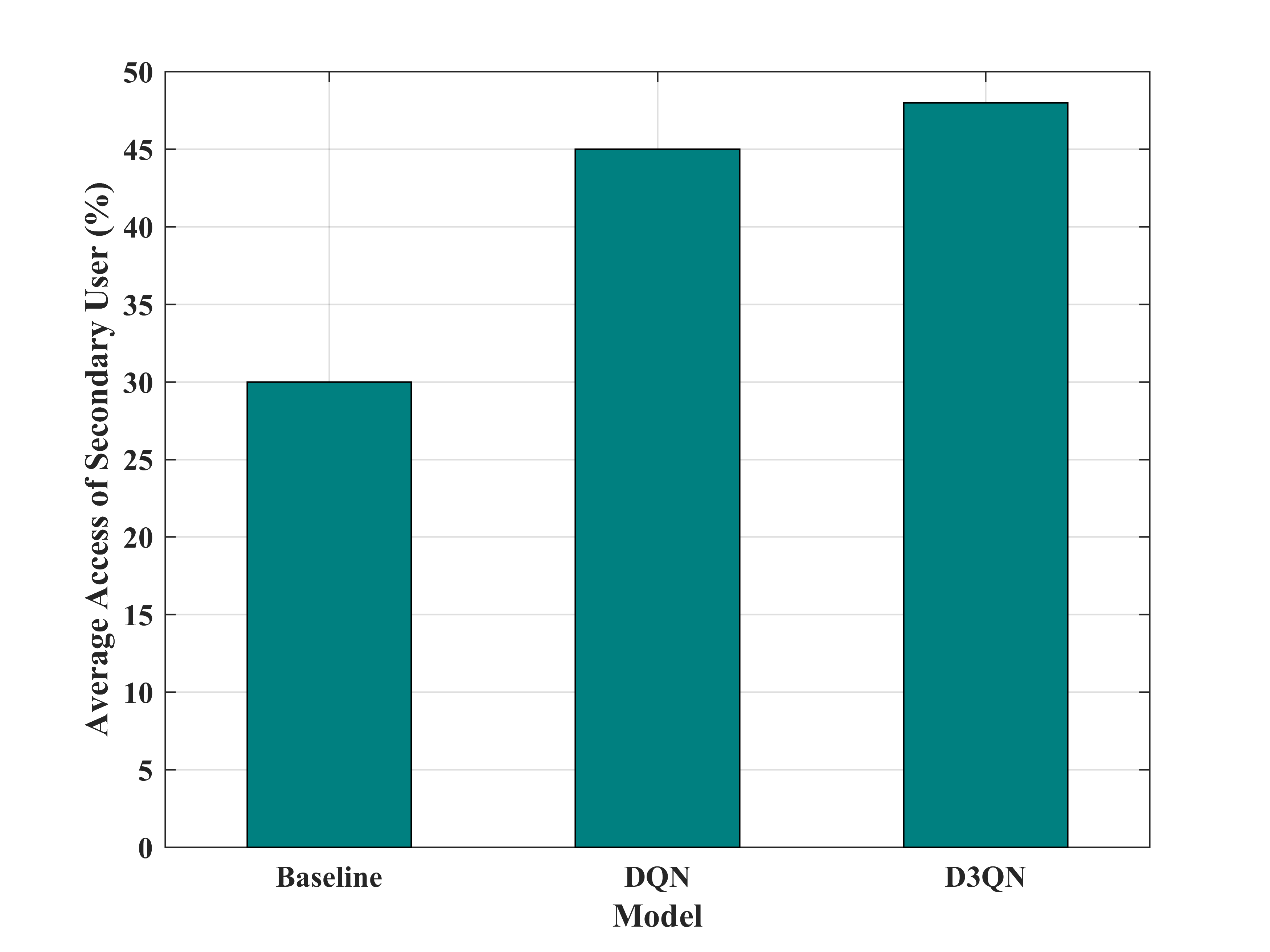}
			\caption{Average user access in different methods}
			\label{Access}
		\end{figure}
		
		The average rate of SU in terms of maximum transmit power of PU is shown in Fig. 7. Because the SU has more access to the spectrum in the proposed case, the average rate is higher than the baseline. On the other hand, with increasing transmit power, the probability of the silent mode occurrence is more than the overlay mode in the baseline. However, in the proposed case, the probability of the underlay mode increases by decreasing the probability of the overlay mode. After that, when the probability of underlay mode decreased, the probability of silent mode increases. According to the figure, at power of 15 dBm and later, because the user is more in the underlay mode, it often starts sending data with the limited power. Therefore, the rate is reduced more than the previous powers, but in the previous powers, the user often sends data with full power during the overlay mode. However, compared to the baseline case, the proposed system has higher rate even at high powers and the D3QN has a slightly higher rate compared to the DQN.
		In Table III, the results of rate and AoI in the proposed DQN and D3QN are compared numerically with the baseline.
		According to the results, the average user access to the spectrum in the baseline case is 30\%, but in the proposed DQN case, it is 45\%, and in D3QN, it is 48\% as shown in Fig. 8. Hence, the reduction in AoI is confirmed.

	\section{CONCLUSION}
		In this paper, intending to minimize the AoI, we considered a novel spectrum sharing system consisting of a primary user and an energy harvesting secondary user in both overlay and underlay modes. The SU sends data with full power in the overlay mode, and it is in the underlay mode with the limited power. The problem was formulated by POMDP and solved with deep Q-network and dueling double deep Q-Network. In this system, the SU must choose an action $x_t$ that reduces the AoI according to the power received from the PU and the amount of energy available in the battery while not interfering with the PU. The simulation results show that use of the proposed system model increases the SU's access time to the spectrum compared to the baseline model, which consists of only overlay mode. In addition, the performance and convergence speed of D3QN is better than that of DQN. Therefore, in the underlay mode, the AoI reduces more than in the overlay mode. Furthermore, considering the normal distribution as the energy distribution performs better than the Poisson distribution and it is considered a more realistic model.


\bibliographystyle{IEEEtran}
\bibliography{References}


\begin{IEEEbiography}[{\includegraphics[width=1in,height=1.25in,clip,keepaspectratio]{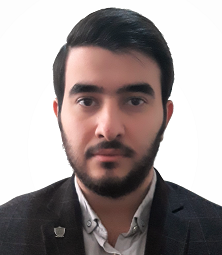}}] {Amir Hossein Zarif}
received the B.Sc. degree in telecommunications engineering from the University of Qom, Qom, Iran, in 2019, and is currently working toward the M.Sc. degree in telecommunications engineering at Tarbiat Modares University, Tehran, Iran. His research interest includes wireless and cellular communications and chipless RFID tag design.
\end{IEEEbiography}
\begin{IEEEbiography}[{\includegraphics[width=1in,height=1.25in,clip,keepaspectratio]{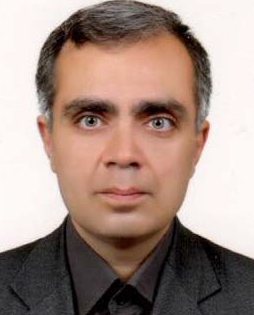}}] {Paeiz Azmi}
was born in Tehran, Iran, in April 1974. He received the B.Sc., M.Sc., and Ph.D. degrees in electrical engineering from the Sharif University of Technology (SUT), Tehran, in 1996, 1998, and 2002, respectively. From 1999 to 2001, he was with the Advanced Communication Science Research Laboratory, Iran Telecommunication Research Center (ITRC), Tehran, where he was with the Signal Processing Research Group, from 2002 to 2005. Since September 2002, he has been with the Electrical and Computer Engineering Department, Tarbiat Modares University, Tehran, where he became an Associate Professor, in January 2006, and is currently a Full Professor. His current research interests include modulation and coding techniques, digital signal processing, wireless communications, radio resource allocation, molecular communications, and estimation and detection theories.
\end{IEEEbiography}
\begin{IEEEbiography}[{\includegraphics[width=1in,height=1.25in,clip,keepaspectratio]{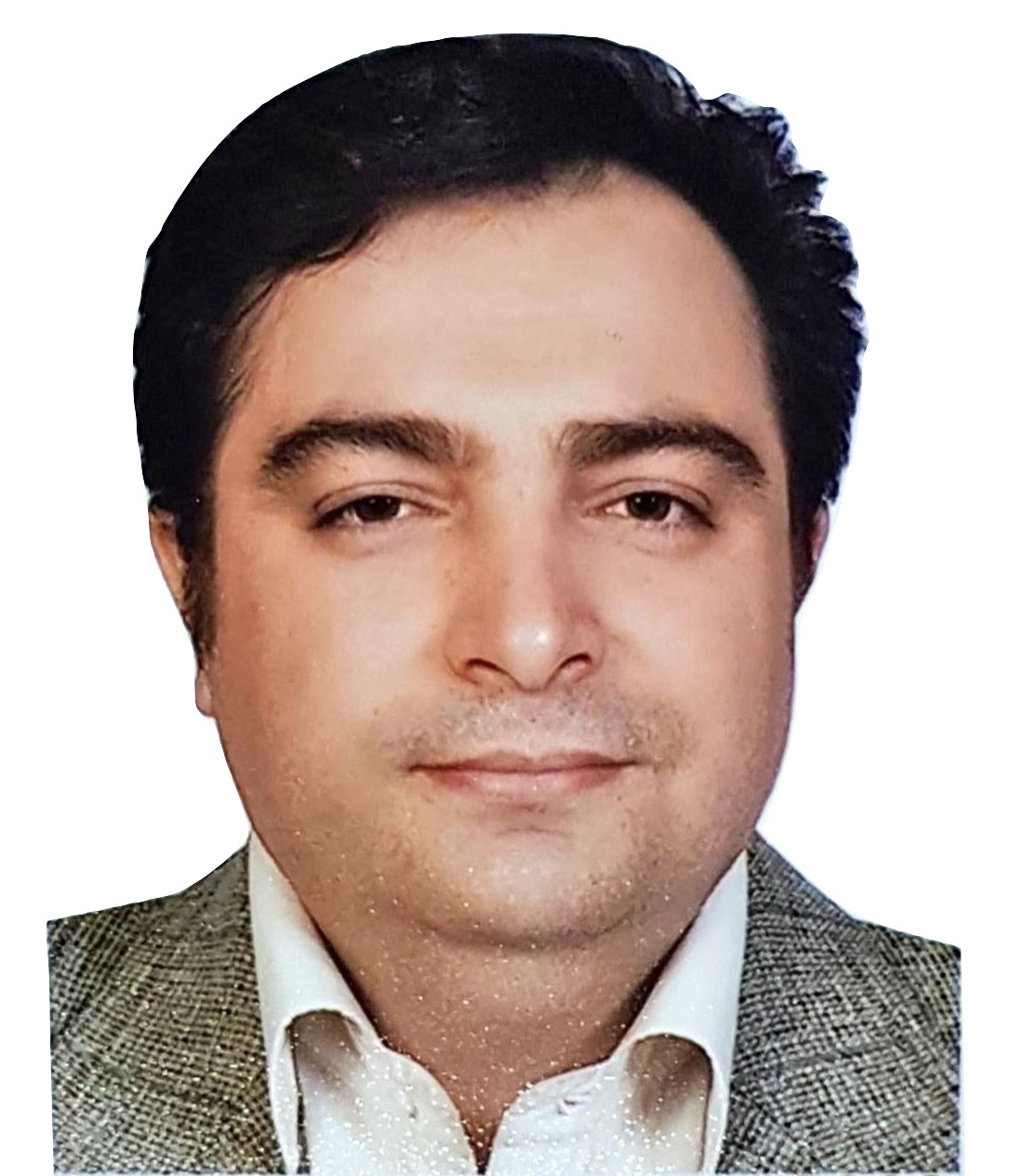}}] {Nader Mokari Yamchi}
completed his PhD studies in electrical Engineering at Tarbiat Modares University, Tehran, Iran in 2014. His thesis received the IEEE outstanding Ph.D. thesis award. He joined the Department of Electrical and Computer Engineering, Tarbiat Modares University as an assistant professor in October 2015. He has been elected as an IEEE exemplary reviewer in 2016 by IEEE Communications Society. Now, he is an Associated Professor at the Department of Electrical and Computer Engineering, Tarbiat Modares University, Tehran, Iran. His research interests cover many aspects of wireless technologies with a special emphasis on wireless networks. In recent years, his research has been funded by Iranian Mobile Telecommunication Companies, Iranian National Science Foundation (INSF). He received the Best Paper Award at ITU K-2020. He was also involved in a number of large scale network design and consulting projects in the telecom industry.
\end{IEEEbiography}
\begin{IEEEbiography}[{\includegraphics[width=1in,height=1.25in,clip,keepaspectratio]{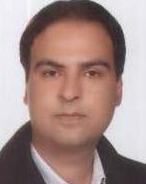}}] {Mohammad Reza Javan}
received the B.Sc. degree from Shahid Beheshti University, Tehran, Iran, the M.Sc. degree from the Sharif University of Technology, Tehran, and the Ph.D. degree from Tarbiat Modares University, Tehran, in 2003, 2006, and 2013, respectively, all in electrical engineering. He is currently a Faculty Member with the Department of Electrical Engineering, Shahrood University, Shahrood, Iran. His research interests include design and analysis of wireless communication networks with emphasis on the application of optimization theory.
\end{IEEEbiography}
\begin{IEEEbiography}[{\includegraphics[width=1in,height=1.25in,clip,keepaspectratio]{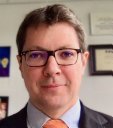}}] {Eduard Jorswieck}
was born in 1975 in Berlin, Germany. He is managing director of the Institute of Communications Technology and the head of the Chair for Communications Systems and Full Professor at Technische Universitaet Braunschweig, Brunswick, Germany. From 2008 until 2019, he was the head of the Chair of Communications Theory and Full Professor at Dresden University of Technology (TUD), Germany. Eduard's main research interests are in the broad area of communications. He has published more than 140 journal papers, 15 book chapters, 3 monographs, and some 280 conference papers on these topics. Dr. Jorswieck is IEEE Fellow. He serves as Editor-in-Chief of the EURASIP Journal on Wireless Communications and Networking. In 2006, he received the IEEE Signal Processing Society Best Paper Award.	
\end{IEEEbiography}
\end{document}